\documentclass[10pt,conference]{IEEEtran}
\usepackage{amssymb}
\usepackage{amsxtra}
\usepackage{amsmath}
\usepackage{latexsym}
\usepackage{graphics}
\usepackage{verbatim}
\usepackage{epsfig}
\usepackage{setspace}

\newtheorem{theorem}{Theorem}[section]
\newtheorem{lemma}[theorem]{Lemma}
\newtheorem{proposition}[theorem]{Proposition}
\newtheorem{corollary}[theorem]{Corollary}

\def\b{\beta}
\def\e{\epsilon}

\def\L{\Lambda}
\def\S{\Sigma}

\def\p{{\mathbf p}}

\def\u{{\mathbf u}}
\def\v{{\mathbf v}}
\def\w{{\mathbf w}}
\def\x{{\mathbf x}}
\def\y{{\mathbf y}}
\def\z{{\mathbf z}}

\def\cD{{\mathcal D}}
\def\cE{{\mathcal E}}
\def\cF{{\mathcal F}}
\def\cG{{\mathcal G}}
\def\cH{{\mathcal H}}
\def\cK{{\mathcal K}}
\def\cL{{\mathcal L}}
\def\cS{{\mathcal S}}
\def\cV{{\mathcal V}}

\def\cX{{\mathcal X}}

\begin{document}

\title{On the Shannon Covers of Certain \\ Irreducible Constrained Systems
of Finite Type$^*$}

\author{
\authorblockN{Akiko Manada and Navin Kashyap}
\authorblockA{Dept.\ Mathematics and Statistics \\
Queen's University \\
Kingston, ON K7L 3N6, Canada. \\
Email: \texttt{\{akiko,nkashyap\}@mast.queensu.ca}}
}
\date{}
\maketitle

\renewcommand{\thefootnote}{$*$}
\footnotetext{This work was supported in part by a research grant 
from the Natural Sciences and Engineering Research Council (NSERC) of Canada.}

\renewcommand{\thefootnote}{\arabic{footnote}}
\setcounter{footnote}{0}

\begin{abstract}
A construction of Crochemore, Mignosi and Restivo in the 
automata theory literature gives a presentation of a finite-type
constrained system (FTCS) that is deterministic and has a 
relatively small number of states. This construction is thus
a good starting point for determining the minimal deterministic 
presentation, known as the Shannon cover, of an FTCS.
We analyze in detail the Crochemore-Mignosi-Restivo (CMR) construction
in the case when the list of forbidden words defining the
FTCS is of size at most two. We show that if the 
FTCS is irreducible, then an irreducible presentation for the 
system can be easily obtained 
from the CMR presentation. 
By studying the follower sets of the states in this irreducible 
presentation, we are able to explicitly determine the Shannon cover
in some cases. In particular, our results show that the CMR construction 
directly yields the Shannon cover in the case of an irreducible 
FTCS with exactly one forbidden word, but this is not in general
the case for FTCS's with two forbidden words.
\end{abstract}


\section{Introduction}

In the information theory literature, constrained systems have 
traditionally arisen in the context of coding for recording systems
\cite{Imm04,HCT,MRS}. 
These systems, under the tag of regular languages, also 
form the cornerstone of automata and formal language theory in 
computer science \cite[Chapters 3--4]{HMU}. 
More recently, constrained systems have come up 
naturally in the context of code design for bio-molecular computation 
(see, for example, the survey paper \cite{BC02}).

To describe the aim of this paper, we need some basic 
terminology \cite{HCT,MRS} from the theory of constrained systems. 
Recall that a {\em labeled graph}, $\cG = (\cV,\cE,\cL)$, is a 
finite directed graph with vertex set $\cV$, edge set 
$\cE \subset \cV \times \cV$, and edge labeling 
$\cL: \cE \rightarrow \Sigma$, where $\Sigma$ is a finite
alphabet. We will refer to the vertices of $\cG$ as \emph{states}.
A \emph{constrained system} (sometimes referred to 
as a \emph{constraint}), $\cS$ or $\cS(\cG)$, is the set of all 
finite-length sequences (\emph{words}) obtained by 
reading off the labels along finite paths in a labeled graph $\cG$. 
The constrained system $\cS(\cG)$
is said to be \emph{presented} by $\cG$; equivalently, $\cG$ is 
called a \emph{presentation} of $\cS(\cG)$. 
A presentation, $\cG$, of a constrained system 
$\cS$ is said to be {\em deterministic} if at each state 
of $\cG$, the outgoing edges are labeled distinctly. 
Deterministic presentations of a constrained system $\cS$ are
used to derive finite-state
encoders for $\cS$ (cf.\ \cite[Chapter~4]{MRS}). 

In general, a given constrained system $\cS$ has many different 
deterministic presentations. However, in practice
it is often desirable to present $\cS$ by a 
deterministic graph with the smallest possible number of states
among all deterministic presentations of the constraint. Such
a \emph{minimal} presentation, called the \emph{Shannon cover}
of the constraint, can be used to find 
finite-state encoders with a small number of 
states which directly translates to low complexity of encoding 
into the constraint. The goal of this paper is to explicitly 
determine the Shannon cover of a certain
class of constrained systems known as irreducible finite-type constraints.

While even the Shannon cover is not in general 
unique for an arbitrary constrained system, it does turn out to be 
unique in the important case of irreducible constrained systems which
we now define. A labeled graph $\cG$ with set of states $\cV$ 
is said to be \emph{irreducible} if for any pair of states $s,t \in \cV$,
there is a directed path in $\cG$ that begins at $s$ and ends at $t$.
A constrained system $\cS$ is defined to be \emph{irreducible} if it can
be presented by an irreducible graph. Equivalently, $\cS$
is irreducible iff for any pair of words $\u, \w \in \cS$, 
there exists $\v \in \cS$ such that the concatenation $\u\v\w$ is also
in $\cS$. It is well known \cite[p.\ 57, Theorem~2.12]{MRS} 
that the Shannon cover of an irreducible constrained system is
unique up to labeled graph isomorphism.

The Shannon cover of an irreducible constrained system $\cS$ can be obtained 
from an irreducible deterministic presentation, $\cG$, of $\cS$ by a procedure
known as \emph{state merging} \cite[Section~2.6]{MRS}. 
This procedure is best described in terms
of the follower sets of states. The \emph{follower set}, $F(s)$, of a state $s$
in $\cG$ is the set of all finite-length words generated by paths in $\cG$
starting at $s$. Two states $s$ and $t$ in $\cG$ 
are said to be \emph{follower-set equivalent} if $F(s) = F(t)$. In such a
situation, states $s$ and $t$ can be \emph{merged} resulting in a 
new graph $\cH$ obtained by first eliminating all edges emanating
from $t$, redirecting into $s$ all remaining edges entering $t$,
and finally eliminating $t$. It is easily verified that the resulting
graph $\cH$ is also an irreducible deterministic presentation of $\cS$.
Recursively carrying out the state merging procedure finally results
in an irreducible deterministic presentation, $\cK$, of $\cS$ that
is also \emph{follower-separated}, which means that distinct pairs
of states in $\cK$ have distinct follower sets. This graph $\cK$
is the Shannon cover of the constraint. In fact, a deterministic
presentation of an irreducible constraint is the Shannon cover
of the constraint iff it is irreducible and follower-separated.

It is clear that the Shannon cover of a constraint $\cS$ is very
simply determined if the state merging procedure can be initiated
on a presentation of $\cS$ that already has 
a relatively small number of states. 
Such a presentation is obtained for a constrained system of finite type
(defined below) via a construction of Crochemore, Mignosi and Restivo 
\cite{CMR98} which has origins in automata theory. 
The Crochemore-Mignosi-Restivo (CMR) construction is 
reasonably amenable to analysis, and we use it as the starting point
in our search for the Shannon cover of a constrained system of finite type.
In fact, Crochemore \emph{et al.}\ prove in a result 
\cite[Theorem~14]{CMR98} related
to the ones in this paper that their construction yields the Shannon cover 
for a certain type of finite-type constrained system. However, our results 
do not follow from theirs.

Let $\S$ be a finite alphabet. We will denote by $\S^*$
the set of all finite-length sequences (words) over $\S$,
including the empty word $\e$. If $\x = x_0 x_1 \ldots x_{\ell-1}$ 
is a word over $\S$, then any of the subsequences 
$x_i x_{i+1} \ldots x_j$, $0 \leq i \leq j < \ell$, is called
a \emph{subword} of $\x$. By convention, the empty word $\e$
is a subword of any $\x \in \S^*$. A \emph{finite-type constrained 
system (FTCS)} is characterized by a finite set $\cF \subset \S^*$, and is
defined to be the set, $\cS_{\cF}$, of all words $\w \in \S^*$
such that $\x$ does not contain as a subword any word in $\cF$. The 
finite set $\cF$ is called a \emph{forbidden set}, and its elements
are called \emph{forbidden words}.
In this paper, we focus mainly on FTCS's
with forbidden sets of cardinality at most two. The difficulties
involved in extending our analysis further 
will already be apparent from the cardinality-two case.

The rest of the paper is organized as follows. The CMR construction is
described in Section~II, and some useful properties of this construction
are given in Section~III. Sections~IV and V study the Shannon covers
of FTCS's with one and two forbidden words, respectively. We show 
there that the CMR construction directly 
yields the Shannon cover in the case of an irreducible FTCS
with exactly one forbidden word, but this is not in general
the case for FTCS's with two forbidden words. Most of the results in
this paper are stated without proof. Complete proofs of these results
will be provided in the full version of the paper.

\section{The CMR Construction}

\begin{figure}[t]
\centerline{\epsfig{file=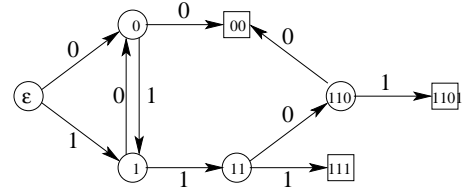, width=6cm}}
\caption{The CMR automaton $\cD_\cF$ for 
$\cF = \{00,1101,111\}$ and alphabet $\S = \{0,1\}$. The squares
represent the sink states.}
\label{cmr_fig1}
\end{figure}

We fix a finite alphabet $\S$, and let $\cF \subset \S^*$ be a 
non-empty finite set. We assume that $\cF$ is a 
\emph{non-redundant} collection of words in that no word $\u \in \cF$
is a subword of any $\w \in \cF$, $\w \neq \u$.
Define a labeled graph $\cD_{\cF} = (\cV,\cE,\cL)$ as follows:
\begin{itemize}
\item $\cV = \{\w : \w \mbox{ is a prefix of a word in } \cF\}$.
Note that a prefix of a word $\x = x_0 x_1 \ldots x_{\ell-1}$
is any of its subwords $x_0 x_1 \ldots x_j$ for $0 \leq j < \ell$,
or the empty word $\e$.  The states corresponding to $\w \in \cF$
will be called \emph{sink states}, and we will often refer to 
the state corresponding to the empty word $\e$ as the \emph{initial state}.
\item There are no edges emanating from any sink state $\w \in \cF$.
There are $|\S|$ edges, all having distinct labels, emanating 
from each state $\u \in \cV \setminus \cF$. These edges are 
defined in the following manner: for each $a \in \S$,
\begin{itemize}
\item[] if $\u a \in \cV$, then the edge labeled $a$ from $\u$
is a \emph{forward edge} that terminates at the state $\u a$; \\[-8pt]
\item[] if $\u a \notin \cV$, the edge labeled $a$ from $\u$
is a \emph{backward edge} that terminates at the state $\v$, 
where $\v$ is the longest suffix 
(incl.\ the empty word $\e$) of $\u a$ in $\cV$.
\end{itemize}
\end{itemize}
The graph thus obtained will be referred to as the 
\emph{CMR automaton} \cite{CMR98}.
Figure~\ref{cmr_fig1} shows such a graph for 
$\cF = \{00,1101,111\}$ and alphabet $\S = \{0,1\}$.

Let $\cG_{\cF}$ be the graph obtained by deleting from $\cD_\cF$
all sink states and all edges entering sink states.
It follows from \cite[Theorem~10]{CMR98} that $\cG_\cF$
is a presentation of the FTCS $\cS_\cF$ having forbidden set $\cF$. 
We will refer to this graph $\cG_\cF$ as the \emph{CMR presentation}
of $\cS_{\cF}$. It is easily seen that
both $\cD_\cF$ and $\cG_\cF$ are deterministic. 
The CMR presentation for $\cF = \{00,1101,111\}$
is the graph given in Figure~\ref{cmr_fig2}, without the dotted edges.

By construction, the number of states in $\cD_\cF$ is at most 
$1 + \sum_{\w \in \cF} \ell(\w)$, and hence, that in $\cG_\cF$ is at most
$1 + \sum_{\w \in \cF} (\ell(\w)-1)$, where $\ell(\w)$ denotes the length 
of the word $\w$. Note that 
$1 + \sum_{\w \in \cF} (\ell(\w)-1) \leq |\cF| \, \ell_{\max}$, where 
$\ell_{\max} = \max \{\ell(\w): \w \in \cF\}$. In comparison, 
the number of states in the canonical
deterministic presentation of $\cS_\cF$ obtained 
from the higher edge graph of
order $\ell_{\max}$ of the unconstrained $\S$-ary system \cite{HCT} 
is $|\S|^{\ell_{\max}-1}$, which is typically much larger than 
$|\cF| \, \ell_{\max}$. Thus, $\cG_\cF$ is in general
a better candidate on which to initiate the state merging procedure to
construct the Shannon cover than the canonical presentation of $\cS_\cF$.

\section{Some Useful Properties of $\cG_\cF$}

In this section, we give some properties of the CMR presentation
$\cG_\cF$ that will be useful in the subsequent development. 
We start with the following observation \cite[Remark~6(1)]{CMR98}, which
is an easy consequence of the definitions of $\cD_\cF$ and $\cG_\cF$. 
\\[-8pt]

\begin{lemma}
For any non-initial state $\u = u_0 u_1 \ldots u_{j-1}$, $j > 0$,
in $\cD_\cF$ or $\cG_\cF$, the edges entering $\u$ all share the same 
label $u_{j-1}$, which is the last symbol of $\u$. 
Hence, a non-initial state in $\cD_\cF$ or $\cG_\cF$ has at 
most one self-loop attached to it.
\label{incoming_labels_lemma}
\end{lemma}
\mbox{} \\[-20pt]

For the CMR automaton $\cD_\cF$, let $\delta: 
(\cV \setminus \cF) \times \S \rightarrow \cV$ be the
\emph{transition function} 
defined by setting $\delta(\u,a)$, $\u \in \cV$, $a \in \S$, to be the state
reached by the edge labeled $a$ emanating from $\u$. Note that 
if the edge labeled $a$ starting at $\u$ is a forward edge, 
then\footnote{Since states in $\cD_\cF$ and $\cG_\cF$ are
identified with words in $\S^*$, the notation $\ell(\u)$ for a state $\u$
simply denotes the length of the word $\u$.}
$\ell(\delta(\u,a)) = \ell(\u a) = \ell(\u) + 1$, and if it is a backward
edge, then $\ell(\delta(\u,a)) \leq \ell(\u)$.

Following \cite{CMR98}, we will find it convenient to define the notion
of a \emph{failure function}, 
$f: \cV \setminus \left(\cF \cup \{\e\}\right) \rightarrow \cV$,
recursively via 
\begin{itemize}
\item for each $a \in \S$, if $\delta(\e,a) \in \cV \setminus \cF$,
then $f(\delta(\e,a)) = \e$;
\item for each $\u \in \cV \setminus \left(\cF \cup \{\e\}\right)$ 
and $a \in \S$, if $\delta(\u,a) \in \cV \setminus \cF$, then 
$f(\delta(\u,a)) = \delta(f(\u),a)$.
\end{itemize}
Note that the failure function is not defined for the initial state and the 
sink states. The usefulness of the failure function stems from the 
fact that it helps in efficiently locating the terminal states 
of the backward edges in $\cD_\cF$ (and hence in $\cG_\cF$). 
Indeed, if for some $\u \in \cV \setminus \left(\cF \cup \{\e\}\right)$ 
and $a \in \S$, we have $\u a \notin \cV$, 
then $\delta(\u,a) = \delta(f(\u),a)$. 

\begin{figure}[t]
\centerline{\epsfig{file=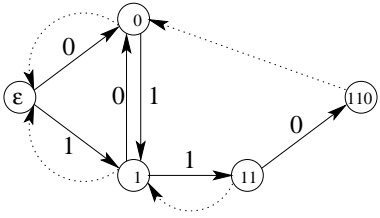, width=5cm}}
\caption{The CMR presentation $\cG_\cF$ for 
$\cF = \{00,1101,111\}$ and alphabet $\S = \{0,1\}$. The dotted
edges represent the failure function.}
\label{cmr_fig2}
\end{figure}

The states in the CMR presentation
$\cG_\cF$ simply retain the failure function (as well as the
transition function whenever it can be defined) from $\cD_\cF$. 
The dotted edges in Figure~\ref{cmr_fig2} represent 
the failure function for the states in $\cG_{\{00,1101,111\}}$.
We will follow this convention of using dotted edges to represent 
the failure function throughout the paper.

We record in  Lemma~\ref{df_lemma}
and Proposition~\ref{f_prop} below 
some facts about the failure function that we will
use in later sections of the paper. 


For a state $\u\neq\e$ in $\cG_\cF$, 
define $\Delta(\u) = \ell(\u) - \ell(f(\u))$. \\[-8pt]

\begin{lemma}
Let $\u,\v$ be non-initial states in $\cG_\cF$ such that $\u$ is a
prefix of $\v$. Then, $\Delta(\u) \leq \Delta(\v)$.
\label{df_lemma}
\end{lemma}
\mbox{}  \\[-20pt]


\begin{proposition}
For any state $\u a$, with $\u \in \S^*$ and $a \in \S$,
in $\cG_\cF$, we have $f(\u a) = \u$ if and only if 
$\u = a^t$ for some integer $t \geq 0$. 
\label{f_prop}
\end{proposition}
\mbox{} \\[-20pt]

\emph{Remark\/}: By convention, $a^0 = \e$. \\[-8pt]


Recall from Section~I that if $\cS_\cF$ is an irreducible constrained
system, then the Shannon cover of $\cS_\cF$ is obtained by applying 
the state merging procedure to an irreducible deterministic presentation
of $\cS_\cF$.  Now, $\cG_\cF$ is certainly deterministic, but need not
always be irreducible. However, it does turn out to be so in
most cases, as we shall see in Sections~IV and V.
The next lemma is a key component in our proofs of irreducibility.


Given a state $\v$ in $\cG_\cF$, let $N_f(\v)$ and $N_b(\v)$ 
respectively denote the number of forward and backward edges that emanate
from $\v$ \emph{in the CMR automaton $\cD_\cF$}. \\[-8pt]

\begin{lemma}
Let $l \geq 0$ be an integer such that every state $\u$ in $\cG_\cF$
with $\ell(\u) \leq l$ has a path leading to a distinguished state $\w$.
If $\v$ is a state with $\ell(\v) = l+1$ such that $N_b(\v) \geq 1$
and either $N_f(f(\v)) < N_b(\v)$ or $\Delta(\v) \geq 2$ holds, then
$\v$ has a path leading to $\w$ as well.
\label{path_lemma}
\end{lemma}
\mbox{} \\[-20pt]


The following result is an application of Lemma~\ref{path_lemma}.

\begin{corollary}
Let $|\S| \geq 3$. If $N_f(\v) \leq (|\S|-1)/2$ for all states
$\v$ in $\cG_\cF$, then $\cG_\cF$ is irreducible.
\label{irred_cor}
\end{corollary}
\mbox{} \\[-20pt]


The last lemma in this section gives an important necessary
condition for two states in $\cG_\cF$ to be follower-set equivalent.
While it only applies to cases in which all forbidden words have the same
length, it is enough for our purposes. 
\\[-8pt]

\begin{lemma}
Let $\cF \subset \S^*$ be a finite set with the property that all words
in $\cF$ have the same length.
If $\x,\y$ are a pair of states in $\cG_\cF$ 
that are follower-set equivalent,
then $\ell(\x) = \ell(\y)$. 
\label{follset_lemma}
\end{lemma}
\mbox{} \\[-30pt]

Note that if $\cF$ consists of exactly one word $\w$, then the 
states of $\cG_\cF$ are precisely all the distinct proper prefixes of $\w$,
which are all of different lengths. We thus have 

\begin{corollary}
If $|\cF| = 1$, then $\cG_\cF$ is follower-separated.
\label{follsep_cor}
\end{corollary}

We investigate the case of forbidden sets of cardinality one in more
detail in the next section.

\section{The Case of One Forbidden Word}

When the forbidden set consists of exactly one forbidden word, 
we have a complete and concise result. \\[-8pt]

\begin{theorem}
Let $\cF = \{\w\}$ for some $\w \in \S^n$, $n \geq 1$. 
If $\cS_\cF$ is irreducible, 
then $\cG_\cF$ is the Shannon cover of $\cS_\cF$.
\label{one_word_thm}
\end{theorem}

\emph{Proof\/}: We have to show that $\cG_\cF$ 
is irreducible and follower-separated. 
By dint of Corollary~\ref{follsep_cor}, it is enough to 
show that $\cG_\cF$ is irreducible whenever $\cS_\cF$ is. In fact,
it is enough to consider the case of a binary alphabet $\S$, since
Corollary~\ref{irred_cor} disposes of the $|\S| \geq 3$ case.

So, let $\S = \{a,b\}$. Without loss of generality (WLOG), 
we may assume that the forbidden word
$\w$ begins with the symbol $a$. 
Note that if $\w = ab^{n-1}$, then $\cS_\cF$ is
not irreducible, since $a,b^{n-1} \in \cS_\cF$, but
there is no word $\x \in {\{a,b\}}^*$ such that $a \x b^{n-1} \in \cS_\cF$.
Similarly, $\cS_\cF$ is not irreducible when $\w = a^{n-1}b$.
For all other words $\w$, as we shall see, $\cG_\cF$ (and hence $\cS_\cF$)
is irreducible.

\begin{figure}[t]
\centerline{\epsfig{file=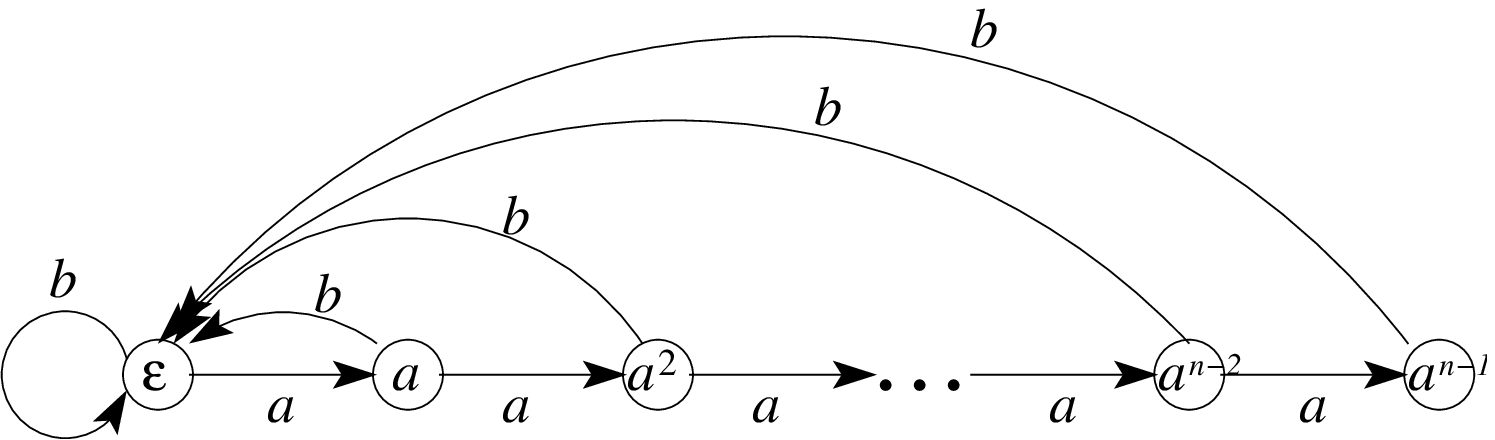, width=6cm}}
\caption{$\cG_\cF$ for $\cF = \{a^n\}$}
\label{thm4.1_fig1}
\end{figure}

We start with $\w = a^n$. It is easily seen that
in this case, $\cG_\cF$ is as in Figure~\ref{thm4.1_fig1}, which
is seen to be irreducible by inspection.

Next, let $\w = ab^ra\y$, with $\y \in \S^*$ and $1 \leq r \leq n-2$.
We will show that all states in $\cG_\cF$ have a path going to the
initial state $\e$.
\begin{figure}[t]
\centerline{\epsfig{file=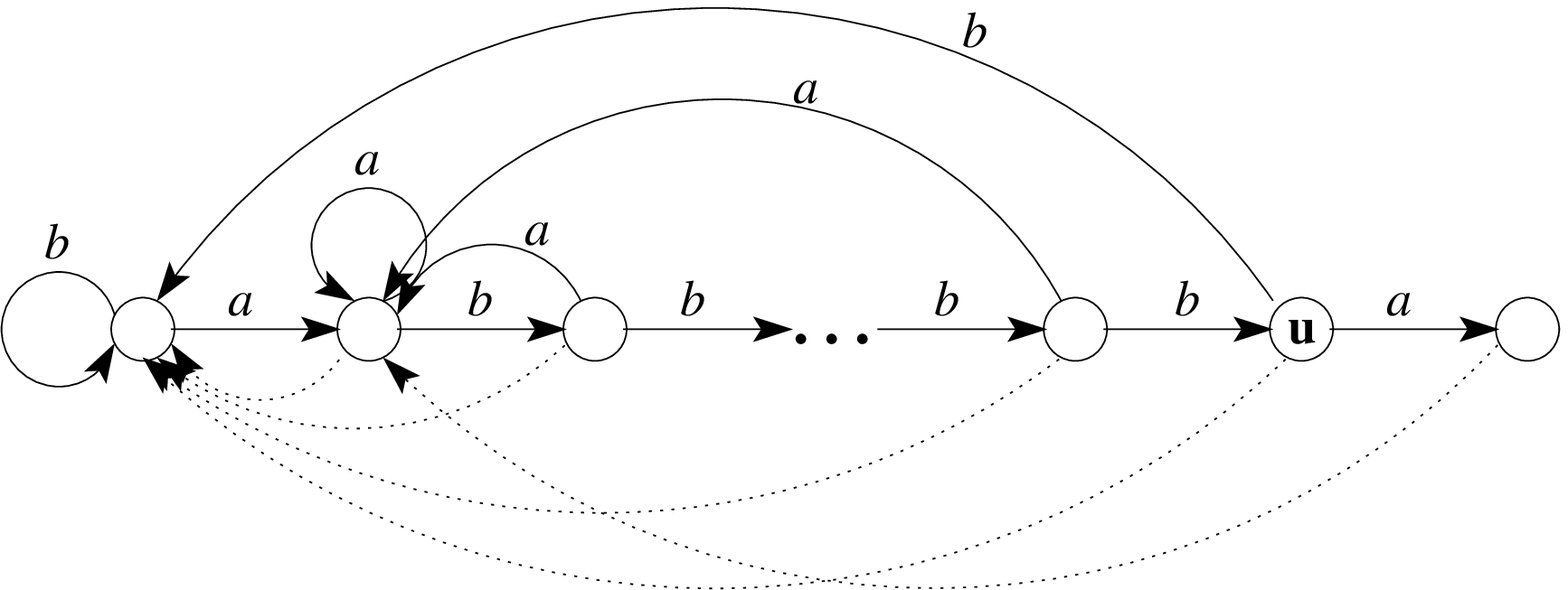, width=6cm}}
\caption{Initial part of $\cG_\cF$ for 
$\cF = \{ab^ra\y\}$, $\y \in \S^*$, $1 \leq r \leq n-2$}
\label{thm4.1_fig2}
\end{figure}
Figure~\ref{thm4.1_fig2} shows the subgraph of $\cG_\cF$ containing
the states from $\e$ up to $\u = ab^r$. Since there is an 
edge from $\u$ to $\e$, we see that there is a path starting
from any state between $\e$ and $\u$ that goes back to the initial state $\e$.
To see that this is also the case for states beyond $\u$, we use
Lemma~\ref{path_lemma}. From Figure~\ref{thm4.1_fig2}, we note 
that $f(\u a) = a$. Thus, $\Delta(\u a) = \ell(\u a) - \ell(f(\u a))
= \ell(\u) \geq 2$. Therefore, by Lemma~\ref{path_lemma}, there is
a path from $\u a$ to the initial state $\e$. For states $\v$ with 
$\ell(\v) > \ell(\u a)$, by Lemma~\ref{df_lemma}, 
we have $\Delta(\v) \geq 2$ as well. So, repeated application of 
Lemma~\ref{path_lemma} shows that any such $\v$ also has a path going
to the initial state $\e$. Thus, $\cG_\cF$ is irreducible.

\begin{figure}[ht]
\centerline{\epsfig{file=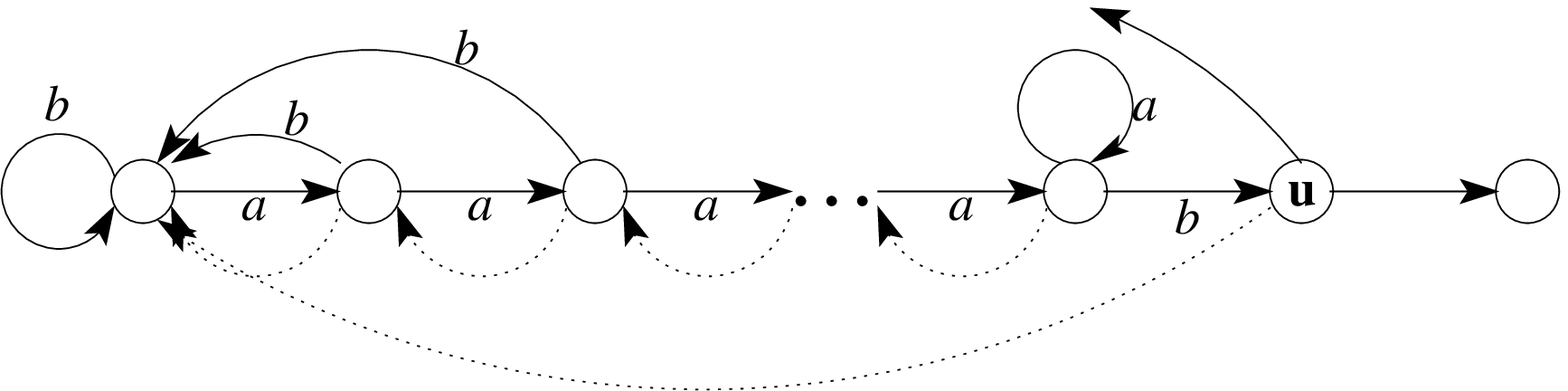, width=6cm}}
\caption{Initial part of $\cG_\cF$ for 
$\cF = \{a^rb\y\}$, $\y \in \S^*$, $2 \leq r \leq n-2$}
\label{thm4.1_fig3}
\end{figure}

Finally, let $\w = a^rb\y$, with $\y \in \S^*$ and $2 \leq r \leq n-2$.
Figure~\ref{thm4.1_fig3} shows the subgraph of $\cG_\cF$ containing
the states from $\e$ up to $\u = a^rb$. We shall show first that there is
a path to the initial state from the state $\u$. Since $f(\u) = \e$,
the backward edge from $\u$, if labeled $a$, goes to the state $a$,
and if labeled $b$, goes to the initial state. But since there is an
edge from the state $a$ to the initial state, there is always a path
from $\u$ to $\e$. In addition since $\Delta(\u) = \ell(\u) \geq 2$,
we also have $\Delta(\v) \geq 2$ for states $\v$ with $\ell(\v) > \ell(\u)$,
by Lemma~\ref{df_lemma}.
Thus, as before, repeated application of 
Lemma~\ref{path_lemma} shows that any such $\v$ also has a path going
to the initial state $\e$, proving that $\cG_\cF$ is irreducible. This
completes the proof of the theorem.   \endproof

\section{The Case of Two Forbidden Words}

When the forbidden set consists of more than one word, the analysis
gets a lot more complicated. The intricacies of the analysis become
evident even in the case of forbidden sets of size two. In this
section, we consider forbidden sets $\cF = \{\w_1,\w_2\} \subset \S^*$,
$\w_1 \neq \w_2$, with $\ell(\w_1) = \ell(\w_2)$. 
Furthermore, we will only present results for the case when $|\S| \geq 3$, 
as the results for the binary alphabet do not have simple 
statements in many cases. For example, when $|\S| \geq 3$, 
$\cG_\cF$ is itself irreducible (Theorem~\ref{two_words_irred_thm}), while 
in the binary case, we sometimes have to pass to a (proper) subgraph of 
$\cG_\cF$ to obtain an irreducible presentation of $\cS_\cF$.

\begin{figure}[t]
\centerline{\epsfig{file=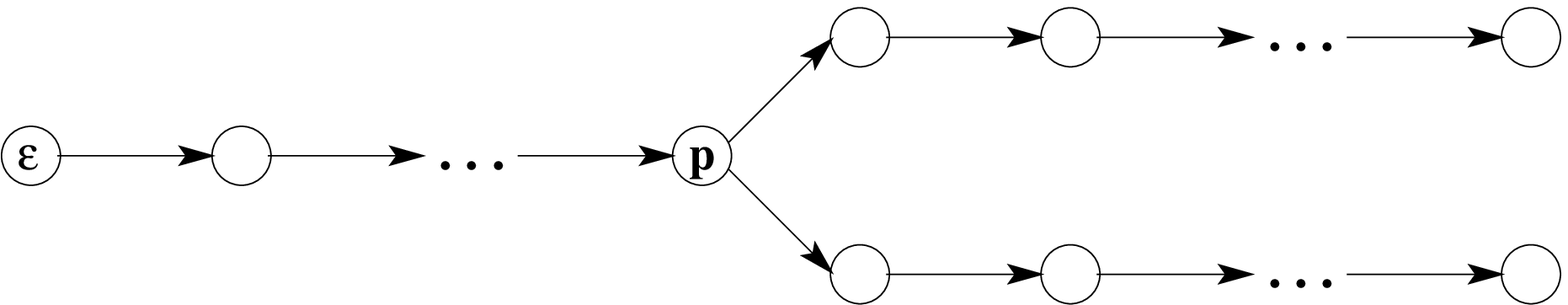, width=7cm}}
\caption{A typical $\cG_\cF$ for $|\cF|=2$. Only forward edges are shown.}
\label{two_word_graph}
\end{figure}

So, for the rest of this section, we will assume a finite alphabet $\S$ with 
$|\S| \geq 3$, and a subset $\cF \subset \S^*$ consisting of two distinct
equal-length words, $\w_1$ and $\w_2$. 
We set $\p$ to be the longest common prefix
(including the empty word $\e$) of $\w_1$ and $\w_2$. Note that by
construction, $\p$ is the only state in $\cG_\cF$ with two 
forward edges; all other states have at most one forward edge.
We say that $\cG_\cF$ \emph{forks} at $\p$, 
as $\cG_\cF$ forks into two branches 
``downstream'' from $\p$, as depicted in Figure~\ref{two_word_graph}.
\\[-8pt]

\begin{theorem}
If $\cS_\cF$ is an irreducible FTCS, then $\cG_\cF$ is irreducible as a 
directed graph.
\label{two_words_irred_thm}
\end{theorem}

\emph{Sketch of proof\/}: 
If $|\S| \geq 5$, then Corollary~\ref{irred_cor} gives
us the irreducibility of $\cG_\cF$. If $|\S| = 4$, then proving
the irreducibility of $\cG_\cF$ is still a relatively easy 
application of Lemma~\ref{path_lemma}. We skip the details.

So, suppose that $\S = \{a,b,c\}$, and let $\ell(\w_1) = \ell(\w_2) = n$,
and $\ell(\p) = \rho$. We will show that 
each state in $\cG_\cF$ has a path leading to the initial state $\e$.
We divide the proof into three cases: (a) $\rho = 0$;
(b) $1 \leq \rho \leq n-2$; and (c) $\rho=n-1$. 
We give complete proofs for the first and last cases 
as illustrations, but skip the proof for Case~(b).

\underline{\emph{Case (a): $\rho = 0$}}. Here, the graph
$\cG_\cF$ forks at the initial state itself. WLOG, the two forward
edges from $\e$ are labeled $a$ and $b$, respectively.
It is enough to show that the states $a$ and $b$ each have a path
going to the initial state. Indeed, if this can be shown, then
it follows from Lemma~\ref{path_lemma} that the states 
$\v$ with $\ell(\v) \geq 2$ also have paths going to the initial state,
as these states satisfy the conditions of that lemma.

Suppose that one of the states $a$ and $b$
has an edge going to the initial state $\e$. WLOG, let this state be $a$.
The state $b$ has two backward edges, of which at most one can be a 
self-loop by Lemma~\ref{incoming_labels_lemma}. 
Thus, the other edge goes either to $\e$ or to $a$. 
In any case, $b$ also has a path going to the initial state. 

\begin{figure}[ht]
\centerline{\epsfig{file=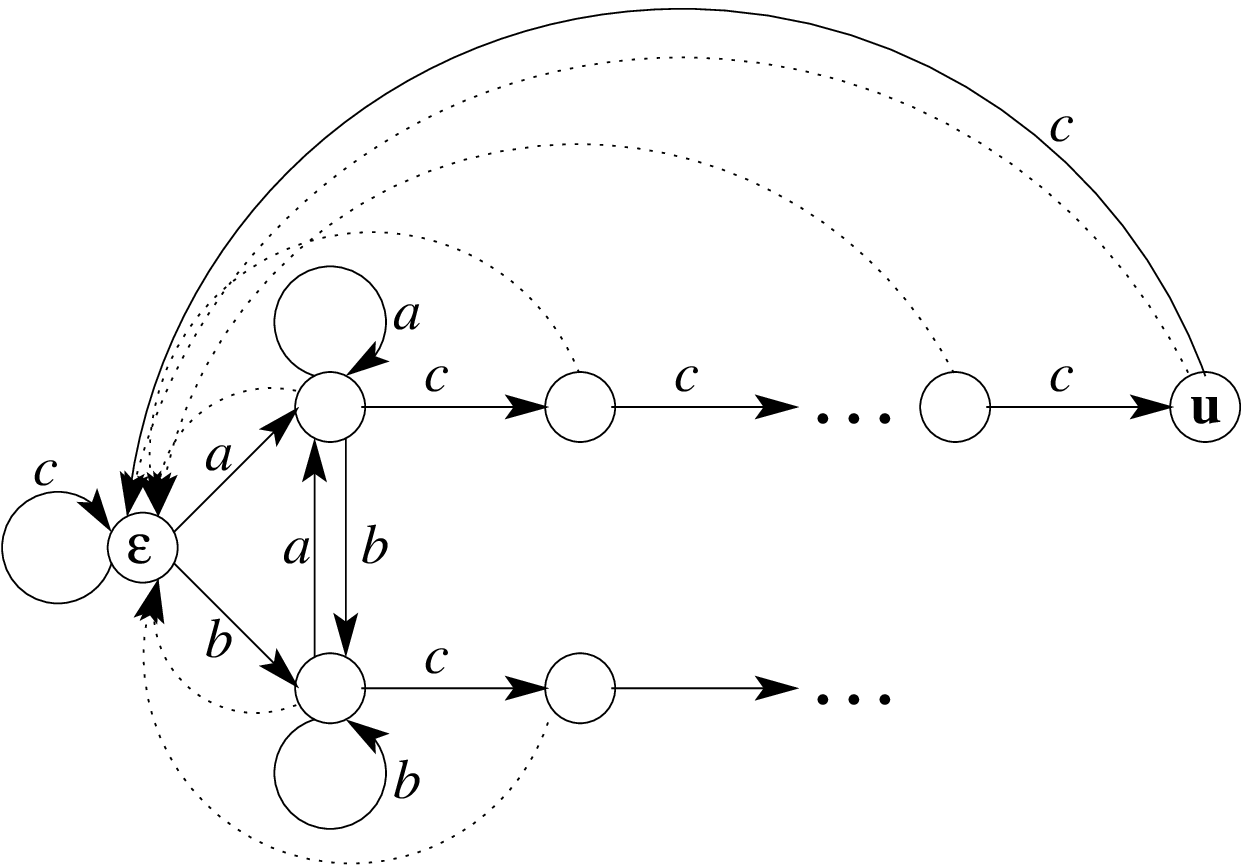, width=6cm}}
\caption{Initial part of $\cG_\cF$ for the case 
$\cF = \{ac\x,bc\y\}$ for some $\x,\y \in \S^*$.}
\label{thm5.1_fig4}
\end{figure}

We are left to deal with the situation when neither $a$ nor $b$ has
an edge going to the initial state. In this situation, 
both $a$ and $b$ have self-loops, $a$ has an edge 
going to $b$, and $b$ has an edge terminating at $a$.
It is straightforward to see that this can happen only if 
$\cF=\{ac\x,bc\y\}$ for some $\x,\y \in \S^*$, 
in which case the initial part of the graph $\cG_\cF$
is as in Figure~\ref{thm5.1_fig4}.

Now, if $\x = \y = c^{n-2}$, then we have $\cF = \{ac^{n-1},bc^{n-1}\}$,
in which case $\cS_\cF$ is not irreducible, 
since $a,c^{n-1} \in \cS_\cF$, but there can
be no $\z \in {\{a,b,c\}}^*$ such that $a\z c^{n-1} \in \cS_\cF$.
So, assuming WLOG that $\x \neq c^{n-2}$, there is a largest $r < n-1$
for which $ac^r$ is a state; let $\u$ denote the state $ac^r$
corresponding to this largest $r$. As shown in Figure~\ref{thm5.1_fig4},
$\delta(\u,c) = \e$. Thus, the states $a$ and $b$ both have paths leading 
to the initial state, as can be verified from the figure.

We have thus proved that $\cG_\cF$ is irreducible whenever $\rho = 0$. 

\begin{figure}[t]
\centerline{\epsfig{file=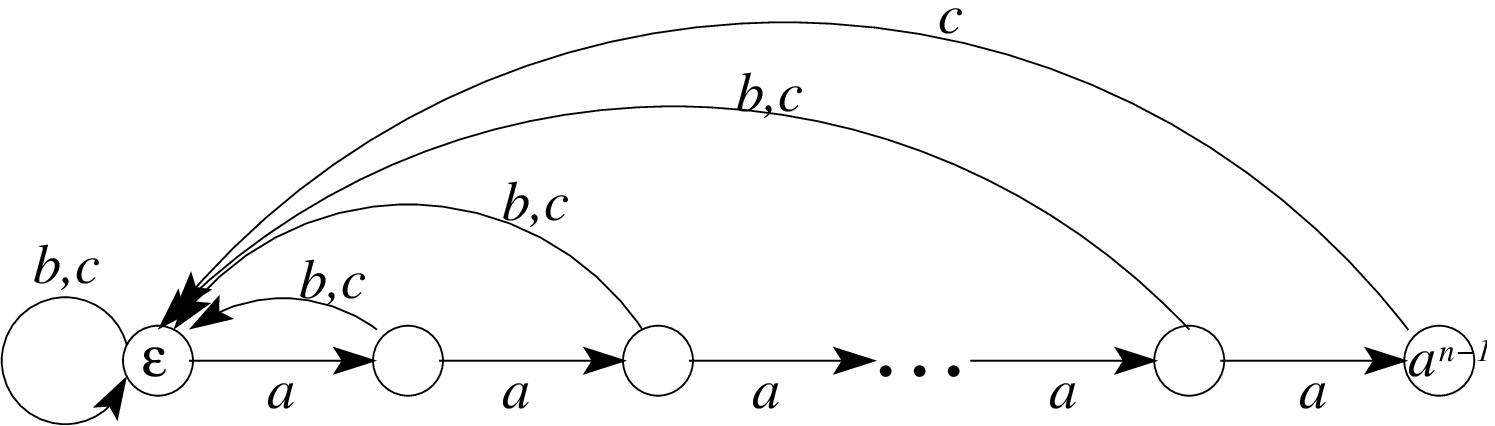, width=6cm}}
\caption{$\cG_\cF$ for $\cF = \{a^n,a^{n-1}b\}$ and $\S = \{a,b,c\}$.}
\label{thm5.1_fig3}
\end{figure}

\underline{\emph{Case (c): $\rho = n-1$}}. 
If $\p \notin \{a^{n-1},b^{n-1},c^{n-1}\}$, then $\cG_\cF$ can easily be shown
to be irreducible using Lemma~\ref{path_lemma}. So, assume WLOG
that $\cF$ is either $\{a^{n-1}b,a^{n-1}c\}$ or $\{a^n,a^{n-1}b\}$. Note, 
however, that when $\cF = \{a^{n-1}b,a^{n-1}c\}$, $\cS_\cF$ is not
irreducible.
When $\cF = \{a^n,a^{n-1}b\}$, $\cG_\cF$ is as shown in 
Figure~\ref{thm5.1_fig3}, and is clearly irreducible. 
\endproof

Thus, by applying the state merging procedure to $\cG_\cF$, we can obtain
the Shannon cover of $\cS_\cF$. To do this, we must of course
identify the states in $\cG_\cF$ that are follower-set equivalent. 
This also turns out to be a non-trivial task, and we are at present able
to give a complete solution only in the special case when 
$\cF = \{a^n, a\x\}$ for some $\x \in \S^{n-1}$, $\x \neq a^{n-1}$.

In the following exposition, we set $\z_1 = a^n$, and 
$\z_2 = a\x$ for some $\x \in \S^{n-1}$, $\x \neq a^{n-1}$. 
Note that we can parse $\z_2$ uniquely as 
$$\z_2 = a^{x_1} \, \b^{(1)} \, a^{x_2} \, \b^{(2)} \, \ldots \,
         a^{x_{q-1}} \, \b^{(q-1)} \, a^{x_q}$$
for some integer $q \geq 2$, where 
$x_1,x_2,\ldots,x_{q-1}$ are positive integers, 
$x_q$ is a non-negative integer, 
and $\b^{(j)} \in {(\S \setminus \{a\})}^*$ for $j = 1,2,\ldots,q-1$.
WLOG, we assume that $\b^{(1)}$ begins with the symbol $b \neq a$.

\begin{figure}[t]
\centerline{\epsfig{file=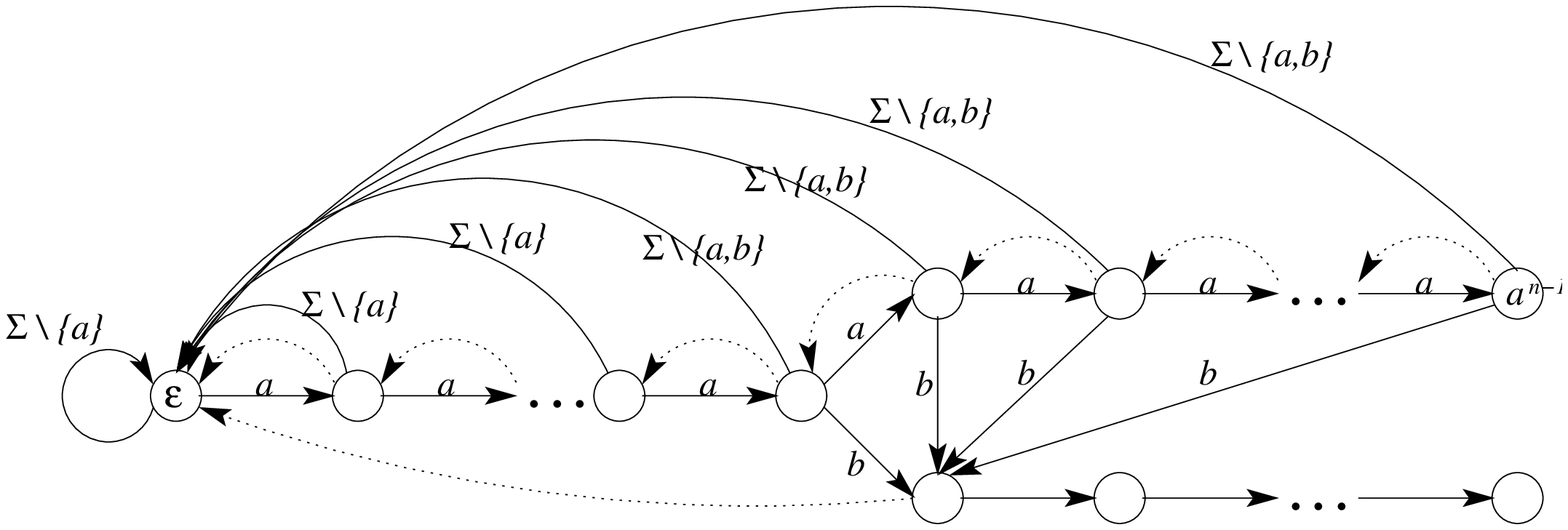, width=7cm}}
\caption{Typical $\cG_\cF$ for the case 
$\cF = \{a^n,a\x\}$ for some $\x \in \S^{n-1}$.}
\label{follsep_fig}
\end{figure}

Figure~\ref{follsep_fig} shows the generic structure of 
$\cG_\cF$ for $\cF = \{\z_1,\z_2\}$. From Theorem~\ref{two_words_irred_thm},
we know that $\cG_\cF$ is irreducible. And as stated in the next
result, this presentation is also follower-separated when $x_1 \geq x_q$.
\\[-8pt]

\begin{theorem}
Let $\cF = \{\z_1,\z_2\}$. If $x_1 \geq x_q$, $\cG_\cF$ is follower-separated,
and hence is the Shannon cover of $\cS_\cF$.
\label{two_words_thm1}
\end{theorem}
\mbox{} \\[-30pt]

To state the corresponding result for the case when $x_1 < x_q$, we need
additional notation and terminology. For $j = 1,2,\ldots,q-1$, we define 
certain distinguished prefixes of $\z_2$,
$$
\p_j = a^{x_1} \, \b^{(1)} \, a^{x_2} \, \b^{(2)} \, \ldots \,
         a^{x_j} \, \b^{(j)} \, a,
$$
and set $\p_0 = a$. The states $\p_j$ in $\cG_\cF$ satisfy the
following property.\\[-8pt]

\begin{lemma}
For $j > 0$, $f(\p_j) = \p_k$ for some $k < j$.
\label{pj_lemma}
\end{lemma} 
\mbox{} \\[-20pt]

Thus, we can define the set of indices
$$
\mbox{Ind}_f = \{k: \ f^r(\p_{q-1}) = \p_k \mbox{ for some } r \geq 1 \},
$$
where $f^r(\cdot)$ denotes the $r$th iterate of the failure function $f$.
Note that $0 \in \mbox{Ind}_f$, since some iterate of the failure
function will eventually take $\p_{q-1}$ to $\p_0=a$.

For $l = 0,1,\ldots,n-1$, let us define
$\L_l$ to be the set of all states $\u$ in $\cG_\cF$ such that 
$\ell(\u) = l$. Thus, $|\L_l| = 1$ if $l \leq x_1$,
and $|\L_l| = 2$ if $l > x_1$.
We will often say that the states in $\L_l$ are 
\emph{at level} $l$ in $\cG_\cF$. Recall from Lemma~\ref{follset_lemma}
that two states in $\cG_\cF$ are follower-set equivalent only if
they are at the same level. We can now state the result for $x_1 < x_q$.
\\[-6pt]

\begin{theorem}
Let $\cF = \{\z_1,\z_2\}$. If $x_1 < x_q$, define
$$
\cX = \{x_{k+1}:\ k \in \mbox{Ind}_f,\ k > 0,
\mbox{ and } x_1 \leq  x_{k+1} < x_q\}.
$$
If $\cX \neq \emptyset$, set $x^* = \max{\cX}$;
else, set $x^* = x_1 - 1$.
Then, the states at level $l \geq \ell(\p_1)$ are
follower-set equivalent iff 
$l \geq \ell(\p_{q-1}) + x^*$. 
Consequently, the Shannon cover of $\cS_\cF$ is obtained from $\cG_\cF$
by merging the pair of states at each level $l \geq \ell(\p_{q-1}) + x^*$.
\label{two_words_thm2}
\end{theorem}
\mbox{}

Theorems~\ref{two_words_thm1} and \ref{two_words_thm2} completely 
specify the Shannon cover in the case of $\cF = \{\z_1,\z_2\}$. As
a direct corollary of these theorems, we have the following result.
\\[-8pt]

\begin{corollary}
For $\cF = \{\z_1,\z_2\}$, the number of states, $\nu_\cF$,
in the Shannon cover of $\cS_\cF$ is given by 
$$
\nu_\cF = \left\{
\begin{array}{cl}
2n - x_1 - 1 & \  \mbox{ if } x_1 \geq x_q \\
2n - x_1 - (x_q-x^*) & \ \mbox{ if } x_1 < x_q
\end{array}
\right.
$$
\label{num_states_z1z2_cor}
\end{corollary}

Generalizing the above result to arbitrary $\cF$'s of size two is by no means
easy. We do have the following simple bound in the case of alphabets of
size at least three, but finding tighter bounds or exact results
remains an open problem. \\[-8pt]

\begin{theorem}
Let $\cF = \{\w_1,\w_2\}$, for some $\w_1,\w_2 \in \S^n$, $\w_1 \neq \w_2$.
Define $\rho$ and $\sigma$
to be the lengths of the longest common prefix and the longest common
suffix, respectively, of $\w_1$ and $\w_2$. Then, the 
number of states, $\nu_\cF$, in the Shannon cover of $\cS_\cF$ 
can be bounded as 
$$
2n - \rho - \sigma - 1 \leq \nu_\cF \leq 2n - \rho - 1
$$
\label{num_states_gen_thm}
\end{theorem}
\mbox{}\\[-30pt]

The results of Theorems~\ref{two_words_irred_thm},
\ref{two_words_thm1}, \ref{two_words_thm2}
and \ref{num_states_gen_thm} can be extended, upon
appropriate modification, to binary alphabets as well. But as the
statements in the binary case are a lot more dense,
we do not present them in this paper. The results for the binary
alphabet, as well as complete proofs of the results given here, will
be published in the full version of this paper.

\end{document}